\documentstyle[preprint,aps]{revtex}
\begin{document}
\draft
\preprint{APL in press}
\title{Coherent Exciton Lasing in ZnSe/ZnCdSe Quantum Wells?}
\author{M.E. Flatt\'e, E. Runge$^*$ and H. Ehrenreich}
\address{Division of Applied Sciences, Harvard University,
Cambridge, Massachusetts 02138}
\date{July 20, 1994}
\maketitle
\begin{abstract}
A new mechanism for exciton lasing in ZnSe/ZnCdSe quantum wells is
proposed. Lasing, occurring below the lowest exciton line, may be
associated with a BCS-like condensed (coherent) exciton state.
This state is most stable at low temperatures for densities
in the transition region
separating the exciton Bose gas and the coherent exciton state.
Calculations show the gain region to lie below the exciton line and to
be separated from the absorption regime by a transparency region of width,
for example, about 80 meV for a 90\AA\  ZnSe/Zn$_{.75}$Cd$_{.25}$ Se
quantum well.
Experimental observation of the transparency region using differential
spectroscopy would confirm this picture.
\end{abstract}

\narrowtext

Time-resolved pump-probe measurements of absorption in
ZnSe/ZnCdSe quantum wells indicate that lasing in these
systems may originate from exciton states\cite{Ding,Ding2,Nurmikko}.
Excitons in this system are unusually robust; the binding energy (40
meV in a 90\AA\ quantum well)
exceeds both $k_BT$ at room-temperature and the
LO-phonon energy (31 meV). This situation sharply contrasts that in GaAs,
where the quantum-well-enhanced binding energy can be
10 meV, and the LO-phonon energy is 36 meV. Carriers in
the ZnSe/ZnCdSe systems relax quickly to quasi-equilibrium distributions
due to  LO-phonon coupling.

Photon emission in this system exhibits two unusual characteristics:
emission occurs at energies below the exciton line and at densities for
which the excitons can no longer be regarded as independent. As a
criterion for the importance of many-body exciton interactions, the
characteristic exciton density  $n_o$ is most simply defined as that
for which the binding energy of an effective exciton in a Bose gas
vanishes. In the presence of an increasing density of other excitons, the
exciton wave functions begins to overlap.  At low temperatures this overlap
marks the
transition to the coherent state described below, while at higher
temperatures it marks the transition to an electron-hole plasma.
Our value
for $n_o$ agrees roughly with that expected on the basis of a Mott
criterion.  Both values are an order of magnitude
smaller than a criterion
used by Nurmikko\cite{Ding,Ding2,Nurmikko} based on densities
when isolated excitons,
characterized by density-independent Bohr orbits, first overlap.
This letter will show how an excitonic picture
survives, with
modifications, at such high densities and how emission occurs below
the exciton line.

The mechanism suggested by Nurmikko {\it et al.} in this
seminal work\cite{Ding,Ding2,Nurmikko} requires an inhomogeneous
exciton
linewidth due to imperfections which substantially exceeds the homogeneous
linewidth.  The exciton states
in their model are non-overlapping and have a
density below the electron-hole plasma transition.
The occupied states lie at the bottom of the inhomogeneous exciton band.
As a result lasing involves excitons lying below the
center of the exciton line.

This mechanism could not be operative in the absence of crystal
imperfections. We propose here a new mechanism for lasing in the
ZnSe/ZnCdSe quantum wells which does not rely on disorder. It is based
on an unusual state of the interacting electron-hole system which
involves a  condensation of electron-hole
pairs\cite{Zimmerman}, similar to the BCS condensation of electron-electron
pairs in a
superconductor. More precisely, at very low, moderate, and higher densities
it corresponds respectively to independent excitons, a Bose condensate
and the BCS state respectively. We shall refer to the last of these as the
``coherent exciton" state. The smooth transition from the bosonic to the
BCS state occurs near the characteristic density $n_o \approx 4 \times
10^{11}$ cm$^{-2}$ in a 90\AA\ ZnSe/Zn$_{.75}$Cd$_{.25}$Se quantum well.

A heuristic sketch of this state is shown in Fig. 1 for the characteristic
density $n_o$. Figure 1a shows the non-interacting electron-hole plasma at
0 K; $\mu _e$ and $\mu _h$ are the electron and hole quasi-Fermi levels
respectively.
This non-equilibrium state is induced by a forward bias across the
active region. The Hamiltonian of this system is

\begin{equation}
H = \sum_{\bf k} \epsilon_c(k) c_{c,{\bf k}}^\dagger c_{c,{\bf k}} +
\sum_{\bf k}\epsilon_v(k) c_{v,{\bf k}}^\dagger c_{v,{\bf k}}
\end{equation}

\noindent
where $c_{c,{\bf k}}^\dagger (c_{v,-{\bf k}})$ creates a conduction
electron (valence hole) with momentum ${\bf k}$.  Spin indices and
sums are implied. Here

\begin{equation}
\epsilon_c(k) = {k^2\over 2m_e}+E_G -\mu_e,\qquad \epsilon_v(k)
= -{k^2\over 2m_h}+\mu_h,\quad  \mu_e+\mu_h =E_G~~~,
\end{equation}

\noindent
and $E_G$, $m_e$ and $m_h$ are the band gap and effective masses
respectively.  In the presence of the electron-hole interaction,

\begin{equation}
H_{e-h} =  - \sum_{{\bf q},{\bf k},{\bf k'}}V(q)
c_{c,{\bf k}+{\bf q}}^\dagger c_{v{\bf k'}}^\dagger c_{v,{\bf k'}
+{\bf q}}c_{c,{\bf k}} ~~~,
\end{equation}

\noindent
a quasi-particle gap forms at the conduction-electron and valence-hole
Fermi surfaces
(shown in Fig. 1b). Here $V(q)$ is the attractive effective electron-hole
Coulomb interaction modified by screening due to the exciton
polarizability. For our calculations
we use a state-independent $V(q)$  whose magnitude and functional form
is determined by the exciton model specified below.

The assumption of singlet pairing between electrons and holes having
momentum {\bf k} and {\bf -k} respectively leads to a gap
equation\cite{Zimmerman}:

\begin{equation}
2\Delta(k) = \sum_q V(q){2\Delta(k+q)\over E_{eh}(k+q)}~~~.\label{gap}
\end{equation}

\noindent
Here

\begin{equation}
E_{eh}(k) = \sqrt{(\epsilon_c(k)-\epsilon_v(k))^2 +
4|\Delta(k)|^2}
\end{equation}

\noindent
and  $\Delta (k)$ is analogous to the gap parameter in the BCS theory.
The number of excitons in a quantum well having unit volume is

\begin{equation}
n = \sum_k \left({1\over
2}-{(\epsilon_c(k)-\epsilon_v(k))\over 2E_{eh}(k)}\right) ~~~.
\end{equation}

The pairing assumption used here is analogous to the pairing assumption in
superconductivity involving spin up electrons of momentum {\bf k} and spin
down electrons of momentum {\bf -k}. The gaps in the
valence and conduction band labelled $E_{eh}/2$ in Fig. 1b each correspond
to half the energy
$E_{eh}$ required to break up the exciton, one half being assigned to the
electron $e$ and hole $h$ respectively.\cite{Footnote}

The sketch in Fig. 1b satisfies the condition $\mu _{eh} = E_G$
defining  the characteristic density $n_o$. The valence-conduction band
gap is seen to be smaller than the corresponding quantity $E_G$ of Fig. 1a
because of band gap renormalization effects associated with the
electron-hole Coulomb interaction. (The corresponding electron-electron and
hole-hole interactions neglected here would further
increase this gap shrinkage).
The gain and absorption regions are illustrated by the arrows marked ``g"
and ``a" respectively. The gaps  $E_{eh}$ produce a transparency regime,
illustrated by the arrow ``tr", separating the two regions. (The
transparency regime width
$2E_{eh}$, whose magnitude is of the order of the exciton binding energy,
should
be experimentally observable if this model is applicable to the present
situation.)  In addition, the enlarged density of states near the gap edge
enhances the gain near the transparency edge. These features are
qualitatively consistent with the observed  lasing spectrum in the
ZnSe/ZnCdSe quantum wells at energies well below the center of the exciton
line\cite{Ding,Ding2,Nurmikko}.

Equation (4) has been solved for $\Delta (k)$ for 30\AA\ and 90\AA\ quantum
wells as well as the idealized two-dimensional case. The results
are shown in Fig. 2.  Since the heavy-hole, light-hole band
splitting is substantial in these systems (79 meV in the 90\AA\
case), the independent sub-band model  and
the rod model, used by Young {\it et. al.}\cite{PYZN} are valid.
These approximations have been used in quantitatively accurate
calculations of
exciton binding energies and optical absorption coefficients in III-V\cite{PHE}
and
II-VI\cite{PYZN} superlattices and quantum wells. In the rod model, the
interaction potential is
approximated
by considering the electron and hole to be rods having length equal to the
well width. Screening effects of $V(q)$ due to the presence of other excitons
are found to be small within this approximation, particularly for the 90\AA\
quantum
well, and will be neglected.

The fundamental absorption coefficient in  the coherent state is given
by\cite{Comte}

\begin{eqnarray}
\alpha(\omega) = {2\pi^2e^2\hbar\over nmc}{2
\over m\hbar\omega}\sum_{\bf k}&&\vert {\bf e}\cdot
{\bf p}_{c,v}({\bf k})\vert ^2\Bigg[-\left({1\over 2}-
{(\epsilon_c(k)-\epsilon_v(k))\over
2E_{eh}(k)}\right)^2\delta(\omega
-\mu_{eh} +E_{eh}(k)) \nonumber\\
&& + \left({1\over 2}+
{(\epsilon_c(k)-\epsilon_v(k))\over
2E_{eh}(k)}\right)^2\delta(\omega
-\mu_{eh} -E_{eh}(k))\Bigg] ~~~.
\end{eqnarray}

\noindent
Here $n$ is the index of refraction, {\bf e} is the photon's
polarization,  and the momentum matrix element
${\bf p}_{c,v}({\bf k})$ is obtained from band-structure calculations using
a ${\bf k\cdot p}$ method\cite{PYZN,PHE}.

Figure 3 shows the gain spectrum for a 90\AA\  ZnCdSe quantum
well laser for three carrier densities as a function of the photon energy
$h\omega$  relative to $\mu_{eh}$. The characteristic density is $n_o = 4
\times 10^{11}$ cm$^{-2}$. The curves have been assigned a linewidth of 10 meV,

corresponding to that measured at low temperatures\cite{Nurmikko}, which is
presumably associated with inhomogeneous broadening. Each curve shows a
sharply peaked gain region and a corresponding absorption region below and
above $\mu_{eh}$ respectively. In a collisionless ideal quantum well these
regions would be separated by a transparency region of width $2E_{eh}$
around $\mu_{eh}$. This feature is somewhat blurred by the phenomenological
linewidth. By contrasting the energy gain at  $3.7\times 10^{11}$
cm$^{-2}$ with that at $4.1\times 10^{11}$ cm$^{-2}$, just above $n_o$, it is
clear that a pronounced increase in gain occurs in the vicinity of
the characteristic density.

The peak is less than an  meV wide and very large in the absence of
broadening effects. For $n > n_o$ and in the absence of broadening, the
gain has a square-root  singularity at the transparency edge. This
divergence is  associated with the  flat-band regions on either side of the
quasi-particle gaps shown in Fig. 1b. Because of the sensitivity of the
peak height  to  broadening, the calculated threshold density is also
sensitive to this effect. The threshold density corresponding to
$100 {\rm cm}^{-1}$ gain, the value cited in Refs. [1-3], is
calculated to be $3.3\times 10^{11}$ cm$^{-2}$,  and lies just 20\% below
$n_o$.
In these experiments the typical excitation density by optical pumping was
$5.\times 10^{11}$ cm$^{-2}$. The experimental
threshold densities were  estimated to be less than
$7.3\times 10^{11}$ cm$^{-2}$, in agreement with the present
estimates\cite{Ding,Ding2,Nurmikko}.

The coherent exciton state may therefore provide an alternative explanation
to the proposed inhomogeneous-linewidth theory of lasing in
ZnSe/ZnCdSe quantum wells. Since this system is the first  lasing
semiconductor having
excitons that are sufficiently robust at ordinary temperatures, these
alternative explanations are both important.
Both viewpoints lead to lasing below the exciton line. The coherent exciton
state becomes important at the characteristic density $n_o$.
Numerical calculations of
the gap parameter characterizing this state yield results that bear
qualitative similarity to experiment. An experimental attempt to measure
the quasi-particle gap with IR radiation, akin to that first used to measure
the
superconducting gap\cite{Tinkham}, may be used to decide the issue.
The experiment in this case, however, is far simpler since the magnitude of
the gap is about 40 meV, which should be observable even in the
presence of broadening effects using differential optical spectroscopy.

We are grateful to J. Ding, B.I. Halperin, P.C. Martin and A. Nurmikko for
helpful discussions.  This work was supported by an ARPA/URI subcontract
through Brown University No. 283-250040 and by the U.S. Advanced Research
Projects Agency (ARPA) through U.S. Office of Naval Research (ONR) Contract
No. N00014-93-1-0549.

\figure{Noninteracting system (a) and interacting system (b)
for the characteristic carrier density $n_o$.  In the
noninteracting case the quasi-Fermi levels of the conduction electrons
and valence holes are $\mu_e$ and $\mu_h$ respectively.
In the interacting case a quasi-particle gap forms at the
conduction-electron and valence-hole Fermi surfaces. The
gaps in the valence and conduction bands labelled $E_{eh}/2$
each correspond to half the energy required to break up the
exciton, one half being assigned to the electron and one
half to the hole. The gain and absorption regions in the
interacting system are indicated by the arrows marked ``g''
and ``a'' respectively. The gaps $E_{eh}$ produce a
transparency regime, indicated by the arrow ``tr''. The
characteristic density is defined as the density where
$\mu_e+\mu_h = E_G$, and therefore the effective binding
energy vanishes.}

\figure{$E_{eh}$ and $2\Delta$ as functions of density for
$90$\AA, $30$\AA, and strictly 2D quantum wells. $2\Delta$
vanishes as the density approaches zero. At the characteristic
density, when the
effective binding energy of the exciton vanishes (at
$4\times 10^{11}$ cm$^{-2}$ for the $90$\AA\ quantum
well), $2\Delta$ has become comparable to the isolated
exciton binding energy.}

\figure{Gain (positive) and absorption (negative) measured
relative to $\mu_{eh}$ for
three densities near the transition to the coherent state:
$2.9\times 10^{11}$ cm$^{-2}$, $3.7\times 10^{11}$ cm$^{-2}$
and $4.1\times 10^{11}$ cm$^{-2}$. A linewidth of $10$meV,
as seen in experiments, has been introduced.  At the lowest density
the gain region is barely detectable. At the middle density the
maximum gain is $200$ cm$^{-1}$. At the highest density,
which is greater than the characteristic density, the gain
has increased even more dramatically. In each case, there is a
transparency region between the gain region and the
absorption region.}

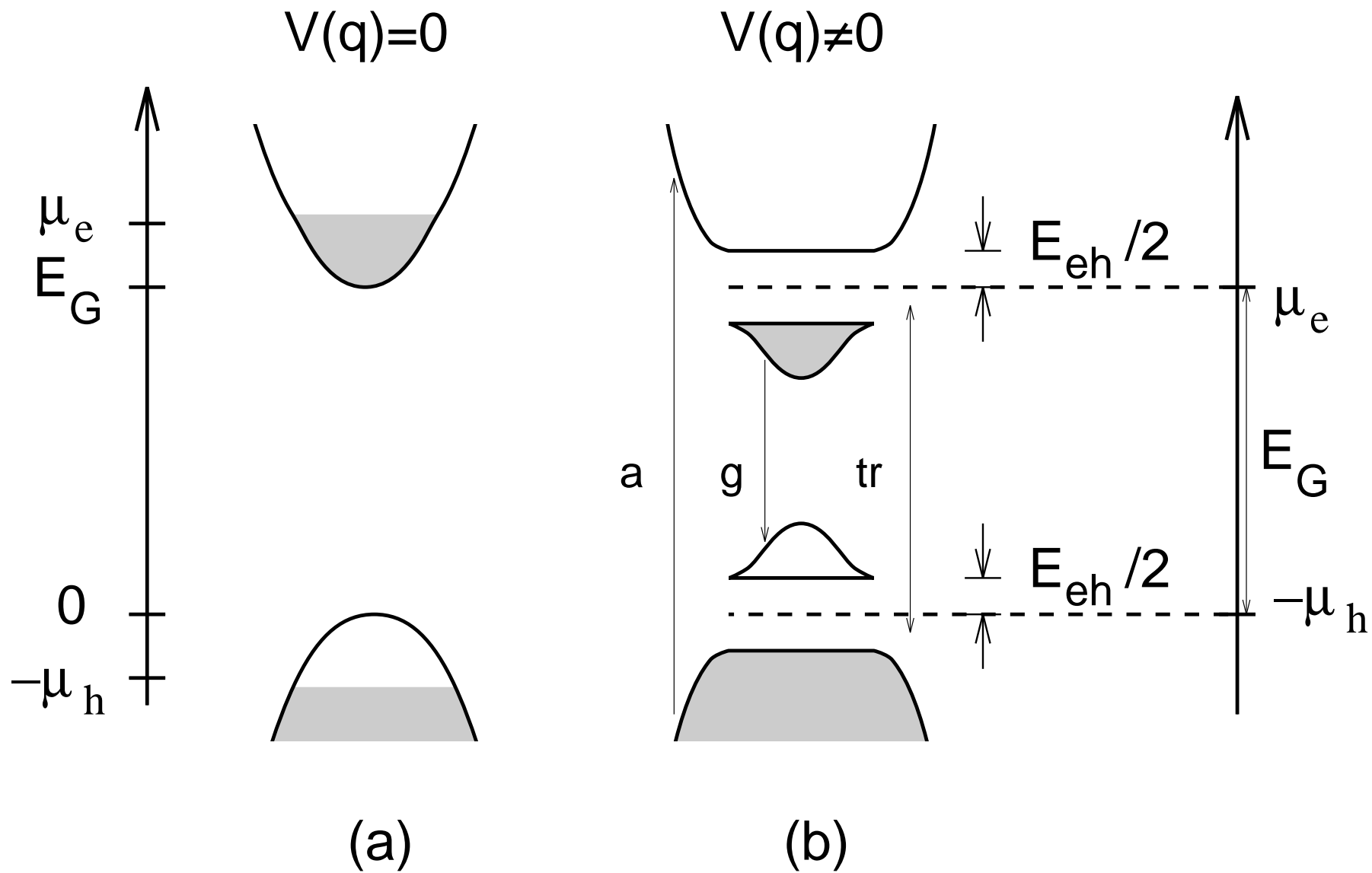
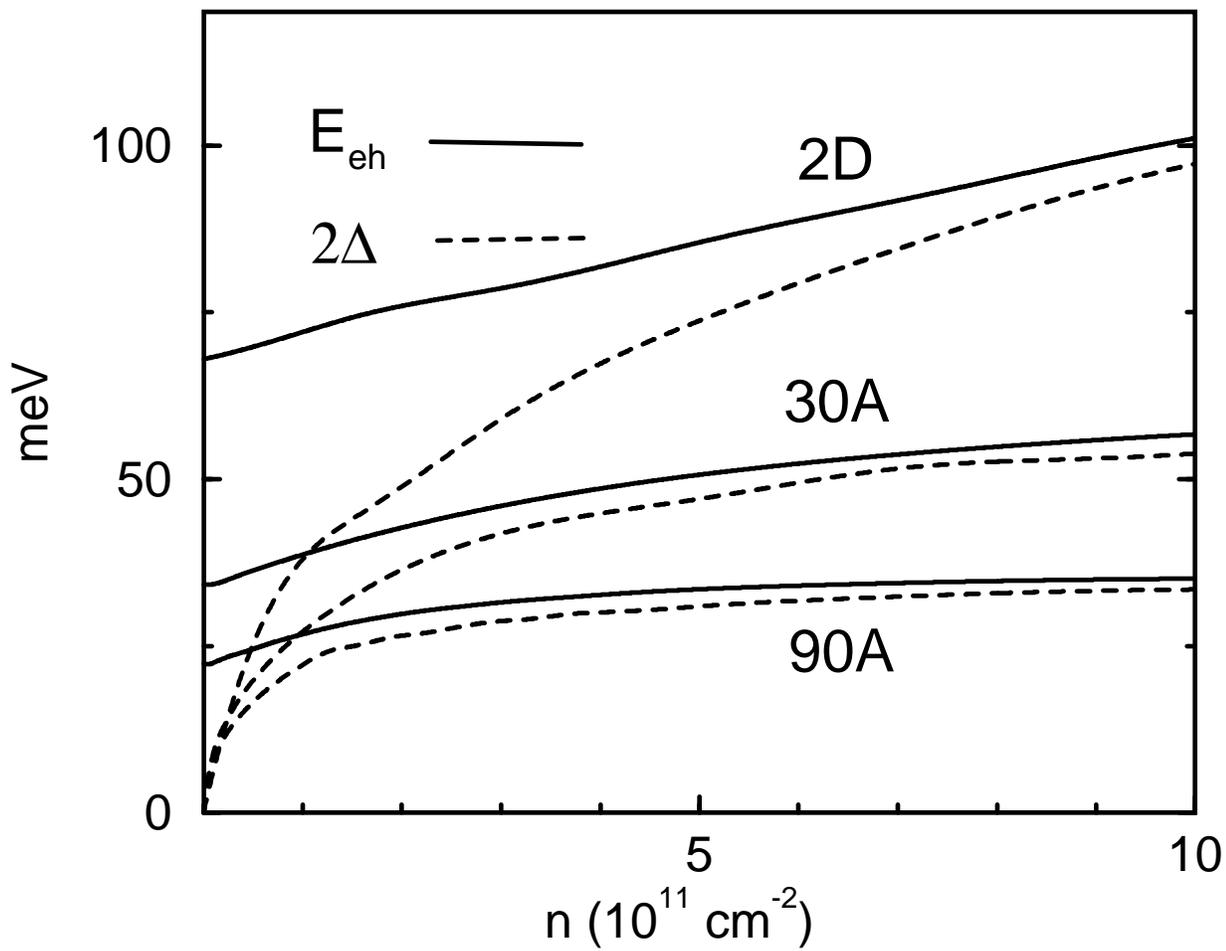
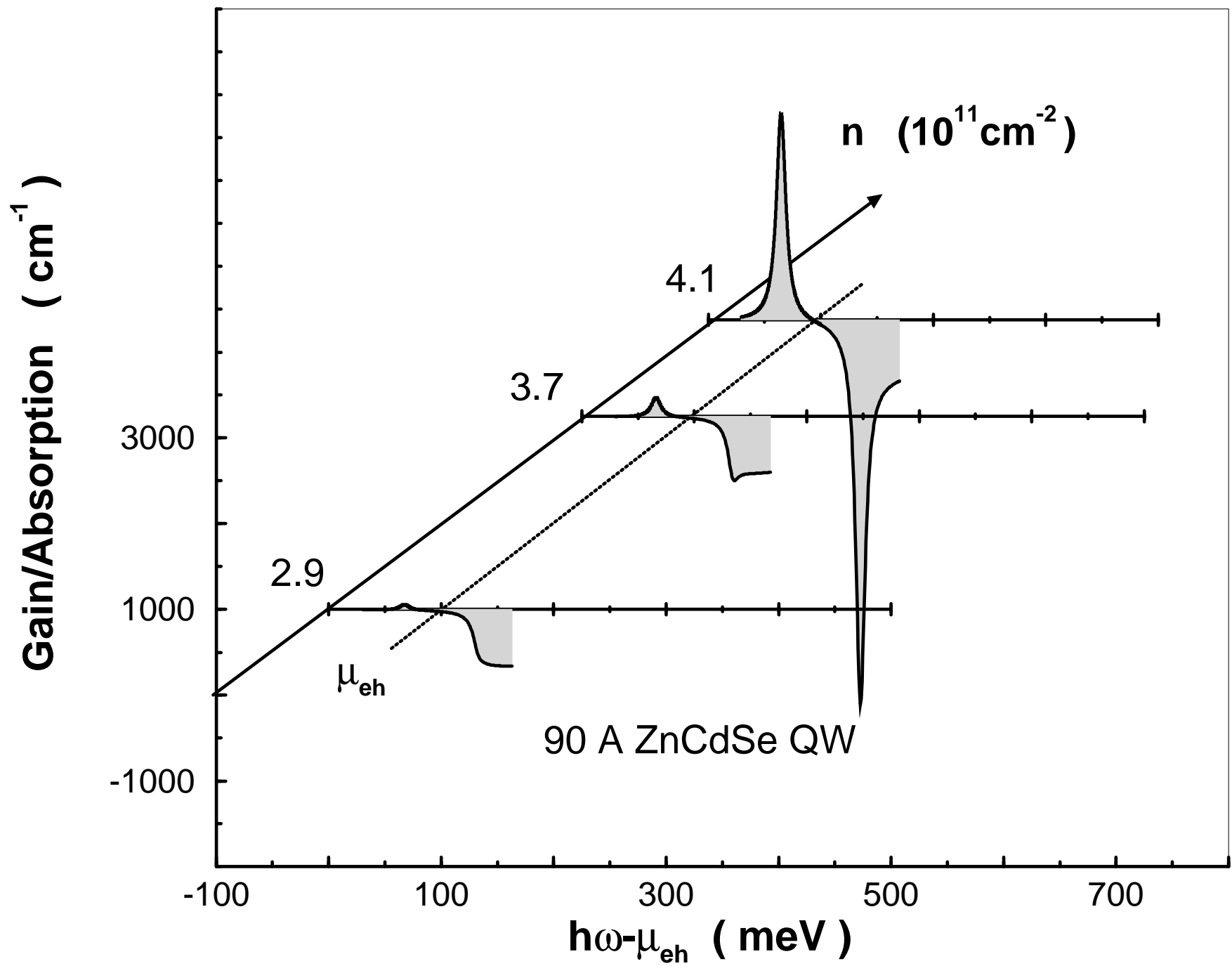
\end{document}